\documentstyle[prb,aps,epsf,twocolumn]{revtex}
\newcommand{\eq}{\begin{equation}}
\newcommand{\ee}{\end{equation}}
\newcommand{\s}{{\sigma}}

\newcommand{\hex}{{\hat{e}}_x}
\newcommand{\hey}{{\hat{e}}_y}

\newcommand{\vrr}{{\vec{r}}}

\newcommand{\vx}{{\vec{x}}}
\newcommand{\vq}{{\vec{q}}}

\newcommand{\vQ}{{\vec{Q}}}

\def\lam{\lambda}
\def\t{\theta}

\def\qt{{\tilde{q}}}
\def\tX{{\tilde{X}}}
\def\tY{{\tilde{Y}}}

\def\e{\epsilon}

\def\ad{a^{\dagger}}

\def\half{{1\over2}}

\def\rhob{{\bar \rho}}

\def\hrho{{\hat \rho}}
\def\ua{\uparrow}
\def\da{\downarrow}
\def\eqa{\begin{eqnarray}}
\def\eea{\end{eqnarray}}
\def\prl{{Phys. Rev. Lett.}}
\def\prb{{Phys. Rev. {\bf B}}}

\def\jpsj{{Jour. Phys. Soc. Japan\ }}

\begin{document}
\draft
\flushbottom
\twocolumn[
\hsize\textwidth\columnwidth\hsize\csname @twocolumnfalse\endcsname
\title{A Composite Fermion Hofstader Problem: Partially Polarized 
Density Wave States in the $\nu=2/5$ FQHE }
\author{  Ganpathy Murthy}
\address{
Physics Department, Boston University, Boston MA 02215\\ and
Department of Physics and Astronomy, Johns Hopkins University,
Baltimore MD 21218}
\date{\today}
\maketitle
\tightenlines
\widetext
\advance\leftskip by 57pt
\advance\rightskip by 57pt

\begin{abstract}
It is well known that the $\nu=2/5$ state is unpolarized at zero
Zeeman energy, while it is fully polarizzed at large Zeeman
energies. A novel state with charge/spin density wave order for
Composite Fermions is proposed to exist at intermediate values of the
Zeeman coupling for $\nu=2/5$. This state has half the maximum
possible polarization, and can be extended to other incompressible
fractions. A Hartree-Fock calculation based on the new approach for
all fractional quantum Hall states developed by R. Shankar and the
author is used to demonstrate the stability of this state to
single-particle excitations, and compute gaps. We compare our results
with a very recent experiment which shows direct evidence for the
existence of such a state, and also with more indirect evidence from
past experiments.

\end{abstract}
\vskip 1cm

]
\narrowtext
\tightenlines
The fractional quantum Hall (FQH) effect\cite{fqhe-ex} has introduced
us to new, highly correlated, incompressible states\cite{laugh} of
electrons in high magnetic fields.  A unified understanding of all
fractions $\nu=p/(2sp+1)$ was achieved by the Composite Fermion
picture\cite{jain-cf}, in which the electrons are dressed by $2s$
units of statistical flux to form Composite Fermions
(CFs). At a mean field level, the CFs  see a reduced field
$B^*=B/(2sp+1)$, in which they fill $p$ CF-Landau levels (CF-LLs), and
exhibit the integer quantum Hall effect. 

Due to the small $g$ factor of electrons in GaAs, spins may not be
fully polarized in FQH states\cite{hal-spin,singlet25}. Transitions
between singlet, partially polarized, and fully polarized states
(based on gap measurements) have been observed for a number of
fillings\cite{clark,initialex,buckthought,duetal}, which can be
understood in terms of CF's with a spin\cite{jain-cf,duetal,wu}. The
transitions happen when an unoccupied CF-LL of one spin crosses the
occupied CF-LL of the opposite spin. Until recently, detailed
measurements of the spin polarization in a FQH state were restricted
to $1/3$\cite{barrett}.

Very recently, Kukushkin {\it et. al} have reported direct
measurements of the total spin polarization for a number of FQH
states\cite{kukush}. In addition to strong transitions
associated with CF-LL crossings, with plateaus of the proper
polarizations predicted by CF theory\cite{wu}, they  observe
weaker plateaus which cannot be explained within the current
framework\cite{duetal}. At $\nu=2/5$ they observe plateaus
corresponding to half the maximal polarization at intermediate values
of the Zeeman coupling\cite{kukush}. These odd features make their
appearance in all the incompressible fractions studied.

I propose a partially polarized density wave (PPDW) state of CFs which
is consistent with the experimental observations. The idea is this: At
$\nu=2/5$ the Zeeman energy $E_Z=g\mu B_{tot}$ favors polarization,
while the electron-electron interaction favors a singlet
state\cite{singlet25,wu}. By forming a periodic structure with two
quanta of {\it effective} flux per unit cell, each CF-LL gets split
into two subbands. The $n=0$ $\da$-spin CF-LL and the $n=1$ $\ua$-spin
CF-LLs can both be half-occupied at intermediate $E_Z$, resulting in
half the maximal polarization. The $n=0$ $\ua$-spin band is always
fully occupied.  When the chemical potential is in the suband gap the
polarization is independent of $E_Z$, leading to a polarization
plateau. Such a state can also have a quantized Hall conductance
(by forming a ``Hall crystal''\cite{hall-crystal}).

Charge density wave (CDW) states of electrons in high magnetic fields
have been explored  in  Hartree-Fock (HF)
theory\cite{yosh-fuku,fuku,yosh-lee,macd}, in which the
interacting problem is converted into a single-electron problem in a
periodic potential, {\it i.e.}, the Hofstader problem. In the strongly
correlated lowest LL (LLL) CDW states of electrons are not attained
until very low densities ($\nu<1/5$)\cite{wc}.

In order to substantiate the idea proposed above, a computational
scheme which gives reliable results for arbitrary FQH states is needed.
Recently R. Shankar and the present author have developed just such an
approach\cite{us1}, based on previous functional
descriptions\cite{func1}.  Our central result is a formula for the
LLL-projected electronic charge density at small $q$:
\eq
\rho_e(q)={\sum_j e^{-iqx_j} \over 2ps+1} -{il_{}^{2} }  (\sum_j (q \times
\Pi_j)e^{-iqx_j}
)\label{rhobar}
\ee
where $\vx_j$ is a CF coordinate, $l=1/\sqrt{eB}$ is the magnetic
length, and ${\vec\Pi}_j={\vec P}_j+e{\vec A}^*(r_j)$ is the velocity
operator of the CFs.  The low-energy Hamiltonian is $ H=\half \int
{d^2 q\over(2\pi)^2} v(q)
\rhob(-q) \rhob(q) $ where $v(q)$ is the electron-electron
interaction. To include the effects of finite sample thickness, and to
stay within the limitations of our small-$q$ approach, we work with a
modified Coulomb interaction of the form $v(q)=e^{-\lam q}2\pi e^2/q$,
where the length $\lam$ is connected to the thickness.  In earlier
papers we presented HF calculations of gaps for a few
fractions\cite{single-part}, tested certain scaling relations that
arise naturally in  our theory against CF-wavefunction
results\cite{scaling}, and computed magnetoexciton
dispersions\cite{me-us} in the TDHF approximation. The results show
that HF in terms of CF variables gives a reasonably good account of
physical properties.

The reason HF is good here is that in our formulation
we are working directly with the quasiparticles of the system, the
CFs. These are quasiparticles in the same sense as in Landau's Fermi
liquid theory: they may interact strongly, but the interaction only
scatters them weakly out of their states at low energies, making a
description in terms of single-particle CF states applicable. Thus we
expect the HF calculation to be described below to be good to the same
level of accuracy (about 20\%). 

We  start with the  wavefunctions for the $n^{th}$ CF-LL
\eq
\phi_{n,X}(\vrr)={e^{{iyX\over l^{*2}}-{(x-X)^2\over2l^{*2}}}\over\sqrt{ l^*L\sqrt{\pi}2^n n!}}
 H_n({x-X\over l^*})
\ee

Here $\vrr$ is a CF coordinate, $L$ is the linear size of the
system (area$=L^2$), and $l^*=l\sqrt{2ps+1}$ is the magnetic
length in the effective field. $X$ is an integer multiple of ${2\pi
l^{*2}\over L}$, and $H_n$ is a Hermite polynomial. The degeneracy of
each CF-LL is ${L^2\over 2\pi l^{*2}}$, which is $1/(2ps+1)$ times the
degeneracy of the electronic LLL.  Using standard manipulations we can
express the momentum space electron density operator as
\eq
\hrho(\vq)=\sum\limits_{n_1,n_2,X}^{} e^{-iq_xX} \ad_{n_1,X_-} a_{n_2,X_+} \rho_{n_1n_2}(\vq)
\ee
where $X_{\pm}=X\pm {q_y l^{*2}\over2}$, $a_{n,X}$ destroys a CF in
the state $\phi_{n,X}$, and the final matrix element is given by
\eqa
&\rho_{n_1n_2}(\vq)={1\over 2p+1}e^{i\t_q(n_1-n_2)-i{\pi\over2}(n_g-n_l)}\sqrt{n_l!\over n_g!} \bigg({ql^*\over\sqrt{2}}\bigg)^{n_g-n_l}\nonumber\\
&\times e^{-{q^2l^{*2}\over4}}(n_gL_{n_l-1}^{n_g-n_l}+2L_{n_l}^{n_g-n_l}-(n_l+1)L_{n_l+1}^{n_g-n_l})
\eea
where $\t_q$ is the angle $\vq$ makes with the $x$-axis, $n_g$ and
$n_l$ are the greater and lesser of $n_1,n_2$, and the $L_n^k$ are the
Laguerre polynomials whose argument is  $q^2l^{*2}/2$.

The Hamiltonian can now be written as 
\eqa
H=&{1\over2L^2}\sum_{\vq\s\s',\{n_i\},\{X_1\}} v(q) e^{-iq_x(X_1-X_2)}
\rho_{n_1n_2}(\vq)\nonumber\\
&\rho_{n_3n_4}(-\vq)\ad_{\s,n_1,X_{1-}}a_{\s,n_2,X_{1+}}\ad_{\s',n_3,X_{2+}}a_{\s',n_4,X_{2-}}\label{ham1}
\eea
where $\s,\s'$ represent the spin labels\cite{implicit}.  The
electronic hamiltonian in the LLL is very similar, except that it must
be normal-ordered to prevent self-interactions. Eq.(\ref{ham1}) above
is the correct form for the CF hamiltonian\cite{us1}, and the energy
coming from normal-ordering represents the Hartree interaction of an
electron with its own correlation hole.

We  allow for a density wave  by making the ansatz\cite{yosh-fuku}
\eq
<\ad_{\s,n_1,X-{Q_yl^{*2}\over2}}a_{\s,n2,X+{Q_yl^{*2}\over2}}>=e^{iQ_xX}\Delta_{\s,n_1n_2}(\vQ)
\ee

In principle, one should allow for nonzero $\Delta$ for all values of
the subscripts, for all $\vQ$'s consistent with the assumed lattice
structure, and repeat for all possible lattice structures. We will
carry out a restricted HF on a square lattice where only
$\Delta_{\s,nn}$ for $n=0,1$ are assumed nonzero, thus ignoring CF-LL
mixing. We consider only the smallest reciprocal vectors $\vQ_1=\pm
Q_0 \hex$, $\pm Q_0
\hey$ with  $\Delta=\Delta_1$, and the next smallest 
vectors $\vQ_2=Q_0(\pm\hex\pm\hey)$, $Q_0(\pm\hex\mp\hey)$ with 
$\Delta=\Delta_2$. It is essential to consider $\vQ_2$ for the square
lattice, because without them there is no gap between the two
subbands, and no partially polarized plateau will be obtained.

I believe no physics is being lost by these restrictions. For
half-filling, the square lattice has a lower HF energy than the
triangular one\cite{yosh-fuku}, and only the above two kinds of
reciprocal vectors need be kept for an accurate
treatment\cite{yosh-fuku}. Also, including (CF-)LL mixing does not
change the physics\cite{macd}.

Since the $n=0$ $\ua$-spin CF-LL is always fully occupied,
$\Delta_{\ua,00}=0$ for all $\vQ$. There are thus
4 variables $\Delta_{1\ua}, \Delta_{2\ua}, \Delta_{1\da},
\Delta_{2,\da}$. The full HF hamiltonian can be written as
\eqa
H_{HF}=&\sum_{\s,n,X} \e^{0}_{\s,n} \ad_{\s,n,X}a_{\s,n,X}\nonumber\\
&+\sum_{\s X,\vQ_i}V_{i\s} \ad_{\s,i,X_+}a_{\s,i,X_-}e^{iQ_xX}
\eea
where, in units of $e^2/\varepsilon l^*$ 
\eqa
&\e^0_{\s,n}=\int\limits_{0}^{\infty}  d\qt {e^{-\lam^*\qt}\over2} \sum\limits_{n'=0}^{n_{max}} (1-2N_F(\s,n'))|\rho_{nn'}(\vq)|^2\\
&V_{i,\s}=\Delta_{i,-\s}{e^{-\lam
Q_i}\rho_{00}(Q_i)\rho_{11}(Q_i)\over Q_il^*}+\nonumber\\
&\Delta_{i,\s}\bigg({e^{-\lam Q_i}\rho_{mm}(Q_i)^2\over Q_il^*}-
\int\limits_{0}^{\infty} d\qt e^{-\lam^*\qt} J_0(\qt Q_il^*)
\rho_{mm}(\qt)^2\bigg)
\eea
where $\qt=ql^*$, $\lam^*=\lam/l^*$, $m=1$ for the $\ua$-spin, and
$m=0$ for the $\da$-spin, and $N_F(\s,n')$ represents the Fermi
occupation of the spin-$\s$ $n'$ CF-LL. In our problem of
interest, $N_F(\ua,0)=1$, $N_F(\ua,1)=N_F(\da,0)=1/2$. 

Now, one diagonalizes the HF-hamiltonian by a sequence of unitary
transformations\cite{yosh-lee} for the case of rational effective flux
through a unit cell. For the square lattice this is
${Q_0^2l^{*2}\over2\pi}={M\over N}$. Each CF-LL splits into $N$
subbands with equal degeneracies. Since we are interested in the
half-filled CF-LL case, we minimally choose $M=1, N=2$, with two
quanta of effective flux through each unit cell. For this simple case
one can write down the energies analytically. They turn out to be
$\e_{\s}=\e^0_{\s}\pm\omega_\s$ where
\eqa
\omega_\s=&2\bigg(V_{1\s}^2(\cos^2{(Q_0X)}+\cos^2{(Q_0Y)})\nonumber\\
&+4V_{2\s}^2\sin^2{(Q_0X)}\sin^2{(Q_0Y)}\bigg)^{\half}
\eea
where $\s=\ua$ implies the $n=1$ CF-LL, and $\s=\da$ the $n=0$ CF-LL.
$X, Y$ are integer multiples of ${2\pi l^{*2}\over L}$ which range
from $0$ to $\sqrt{\pi}$. Finally we impose the self-consistency
conditions
\eqa
\Delta_{1\s}=&-{V_{1\s}\over\pi Q_0^2}\int\limits_{0}^{\pi} d\tX d\tY {\cos^2{(\tX)}\over \omega_s}\nonumber\\
\Delta_{2\s}=&-{2V_{2\s}\over\pi Q_0^2}\int\limits_{0}^{\pi} d\tX d\tY {\sin^2{(\tX)\sin^2{(\tY)}}\over \omega_s}\label{selfcons}
\eea

The gap between the two subbands is $8|V_{2\s}|$. Let us denote the
top of the occupied subband as the highest occupied molecular orbital
(HOMO), and the bottom of the unoccupied subband as the lowest
unoccupied molecular orbital (LUMO). In the absence of Zeeman energy I
find that $E^{\ua}_{HOMO}>E^{\da}_{LUMO}$, which 
is inconsistent; eq.(\ref{selfcons}) is predicated on the chemical
potential lying in both the subband gaps, which must therefore
intersect.  As the Zeeman energy increases the entire $n=0$ $\da$-spin
structure moves up, while the $n=1$ $\ua$-spin structure moves
down. There is a lower critical Zeeman energy $E_{Z}^{<}$ for which
the two subband gaps intersect. As the Zeeman energy increases
further eventually we will have an upper critical Zeeman energy
$E_{Z}^{>}$ where $E^{\ua}_{LUMO}=E^{\da}_{HOMO}$. For
$E_{Z}>E_{Z}^{>}$ the proposed state is again inconsistent. The PPDW
state exists and is stable (to single-particle excitations) for
$E_{Z}^<<E_Z<E_{Z}^>$.

Now for the numerical results. I have found self-consistent solutions
for $\lam=l, 1.5l$, since these values are likely to bracket the
thickness of physical samples\cite{morf-comment}. I find that the PPDW state is
locally stable for $0.006<E_Z<0.012$ at $\lam=l$, while for
$\lam=1.5l$ the regime of stability is $0.002<E_Z<.01$. Here and below
$E_Z$ and all gaps are  in units of $e^2/\varepsilon l$.

The strongest experimental evidence for PPDW states comes from the
total polarization measurements of Kukushkin {\it et
al}\cite{kukush}. Examining their curve for $\nu=2/5$ we see that the
partially polarized state with a constant polarization of half the
maximum value is seen for $0.009<E_Z<0.011$. This is fully consistent
with our results for $\lam=l$, since all we have ascertained is the
stability of the PPDW state to single-particle
excitations, and ignored other instabilities. 

Evidence for PPDW states is seen in their data for other
incompressible fractions as well\cite{kukush}. For example, their
curve for $\nu=2/3$ shows a small region of polarization $1/2$, which
is impossible to explain using translationally invariant states of
CFs. Similar regions are seen in the $4/7$, $3/7$, and $4/9$
data. This leads us to believe that the PPDW state is a generic
feature of the incompressible fractions where the ground state is not
fully polarized at $E_Z=0$. While the polarization measurement is
unambiguous, it would be nice to obtain direct evidence of the
inhomogeneity, perhaps by surface acoustic wave measurements\cite{willett}. The period for
the square lattice is $2l^*\sqrt{\pi}=7.93l$.

There is also suggestive evidence from earlier experiments. For $4/3$,
$8/5$, and $7/5$, Du {\it et al}\cite{duetal} find that while the gap
seems to go to zero at the transition, it does not come back up to the
expected value, but remains (for the latter two fractions) an order of
magnitude below. Theoretically, gaps in the PPDW state are subband
gaps, and are naturally small. For example, at $\lam=l$ in our
approach, the singlet state has a gap of 0.045, the fully polarized
state has a gap 0.032, while the partially polarized state gap (which
changes with $E_Z$) has a maximum value of 0.0029, an order of
magnitude below the usual FQHE gaps. An earlier study\cite{clark}
found a ``gapless'' region at intermediate $E_Z$ for $4/3$ beyond
which the gap recovered. Presumably the gap was below the resolution
in that experiment, since a later experiment\cite{buckthought} found
quantization of $\s_{xy}$ in the ``gapless'' region. Du {\it et al}
also find maxima in the longitudinal resistance not only at the CF-LL
level crossings (the strong maxima), but also at other values of $E_Z$
(which are weaker maxima), which they are unable to
understand\cite{duetal}. These weaker maxima may be caused by
transitions to PPDW states.

Finally, there is some evidence from exact diagonalizations of small
systems\cite{chakraborty} that the $3/5$ and $4/9$ states have a
``gapless'' regime at intermediate $E_Z$. The period of the PPDW state
is too large to show up in these studies.

Let us turn to some of the caveats concerning our calculation, where
many effects are ignored, including CF-LL mixing and larger
$\vQ$'s. We have concentrated on 2 units of effective flux per unit
cell. This is symmetric and natural in this problem since we need the
$\ua$-spin CF-LL with filling $\nu_1$, and the $\da$-spin CF-LL with
filling $1-\nu_1$ to have large gaps. This is also what is seen in the
data\cite{kukush}. However, other configurations are possible. We have
treated the constraint by cutting off the number of CF-LLs
($n_{max}=4$) as in our previous
calculations\cite{single-part,me-us}. As the fractions approach a
compressible state, a better treatment of the constraint will become
necessary\cite{comment}. We have used the small-$q$ version of the
density, whereas the actual value is $Q_0\approx l^{-1}$. This can be
remedied by using a recent extention of our formulation\cite{aftermath} to all
$q$. If a charge density inhomogeneity exists, the
CF's will feel a periodic Chern-Simons field due to the charge
texture. The present treatment produces weak charge 
modulation ($\rho(Q_0)/\rho(0)\approx0.01$), and we have ignored this
effect\cite{iye}. We have only considered the stability to single particle
excitations, while there may be other instabilities as well. One
omission is significant: We have not compared ground state energies of
the various states. The reason is that the ground state structure
factor $S(k)$ in our approach is not correct (at small
$k$)\cite{GMP}. Our approach is an expansion in gradient
interactions\cite{us1}. Due to the peculiar properties of the
electronic density matrix elements, the order we have kept suffices
for reasonably accurate calculations of gaps, but one needs to go to
higher order to get $S(k)$ accurately\cite{us1}.

Let us now explore some potential generalizations of our results. An
obvious extention is to the $\nu=2$ state. My preliminary results
indicate that the PPDW state should be locally stable here as
well\cite{future}.  To access this region experimentally would require
huge in-plane fields, or very low densities.  A loose analogy can also
be made between the $2/5$ two CF-LL problem and the bilayer quantum
Hall problem\cite{bilayer}, with the CF-LL index playing the role of
the layer index. The two fully filled CF-LLs in $\nu=2/5$ correspond
to the $\nu=2$ bilayer problem, for which a translationally invariant
canted antiferromagnetic state has been proposed\cite{canted} at
intermediate $E_Z$. A general 3-parameter wavefunction has been
written for such states\cite{mrj}, which should compete with the PPDW
state.  (For $\nu=2/5$ in our formalism, the analog of this
state\cite{mrj} seems to have no HF solutions other than the singlet
and fully polarized states\cite{future}: In any case, the canted
states should show a smooth variation of the total polarization with
$E_Z$, not the plateaus seen in the data\cite{kukush}). If we push the
analogy further, the PPDW state should occur in the bilayer $\nu=2$
case for the two layers having unequal fillings $1/2$ and $3/2$. These
generalizations will be explored elsewhere\cite{future}.

In summary, I have presented theoretical arguments in support of a
partially polarized density wave state of Composite Fermions at
$\nu=2/5$ (and other incompressible fractions), which can be accessed
by tuning the Zeeman energy. I have also argued that such states have
already been seen in a recent experiment\cite{kukush}, and indicated
suggestive features of older experiments which hint at the existence
of these states. It would be fascinating to explore the connections to
other inhomogeneous states proposed in the quantum Hall
regime\cite{hall-crystal,brey,koulakov}.

It is a pleasure to thank H.A.Fertig, J.K.Jain, and Z.Tesanovic for
illuminating conversations.


\begin{thebibliography}{100}
\bibitem{implicit} We are implicitly attaching flux independent of spin\cite{me-us}.
\end{thebibliography}
\end{document}